\documentclass{iaus}
\usepackage{graphicx}

\newcommand{\MoH}{\ensuremath{\left[\mathrm{M}/\mathrm{H}\right]}}

\title[Photometric colors of late-type giants] 
{Photometric colors of late-type giants: theory versus observations}

\author[Ku\v{c}inskas {\it et al.}]   
{A. Ku\v{c}inskas$^1$$^2$, P.H. Hauschildt$^3$, H.-G. Ludwig$^4$, I. Brott$^3$, \break
 V. Vansevi\v{c}ius$^5$, L. Lindegren$^6$, T. Tanab\'{e}$^7$ \and F. Allard$^8$}

\affiliation
{
$^1$National Astronomical Observatory of Japan, Mitaka, Tokyo 181-8588, Japan
    \break email: arunas.kucinskas@nao.ac.jp\\[\affilskip]
$^2$Institute of Theoretical Physics and Astronomy, Go\v {s}tauto 12, Vilnius 01108, Lithuania
    \break email: ak@itpa.lt\\[\affilskip]
$^3$Hamburger Sternwarte, Gojenbergsweg 112, 21029 Hamburg, Germany\\[\affilskip]
$^4$GEPI - CIFIST, Observatoire de Paris-Meudon, 5 place Jules Janssen, \break 92195 Meudon Cedex , France\\[\affilskip]
$^5$Institute of Physics, Savanoriu 231, Vilnius 02300, Lithuania\\[\affilskip]
$^6$Lund Observatory, Lund University, Box 43, SE-221 00 Lund, Sweden\\[\affilskip]
$^7$Institute of Astronomy, The University of Tokyo, Mitaka, Tokyo, 181-0015, Japan\\[\affilskip]
$^8$CRAL, \'{E}cole Normale Sup\'erieure, Lyon, Cedex 07, 69364 France
}

\pubyear{2005}
\volume{232}  
\pagerange{1--6}
\date{}
\jname{The Scientific Requirements for Extremely Large Telescopes}
\editors{P. Whitelock, B. Leibundgut \& M. Dennefeld, eds.}
\begin{document}

\maketitle

Late-type giants (i.e., stars on the red and asymptotic giant
branches, RGB/AGB, respectively) are dominant contributors to the
overall spectral appearance of intermediate age and old stellar
populations, especially in the red/near-infrared part of the
spectrum. Being intrinsically bright, they are well suited for
probing distant/obscured populations, especially those that can
not be studied with their fainter members, like main sequence turn-off
stars or subgiants. Late-type giants and supergiants will be
the only stellar types accessible in intermediate age and old
populations beyond the distances of several Mpc with the future
30-50\,m class extremely large telescopes (\cite[Olsen \etal\
2003]{O06}). Indeed, proper understanding of their observable
properties by means of theoretical models is of key importance for
studying the evolution of stellar populations and their host galaxies.

\begin{figure}
\centering
\includegraphics[width=13.3cm] {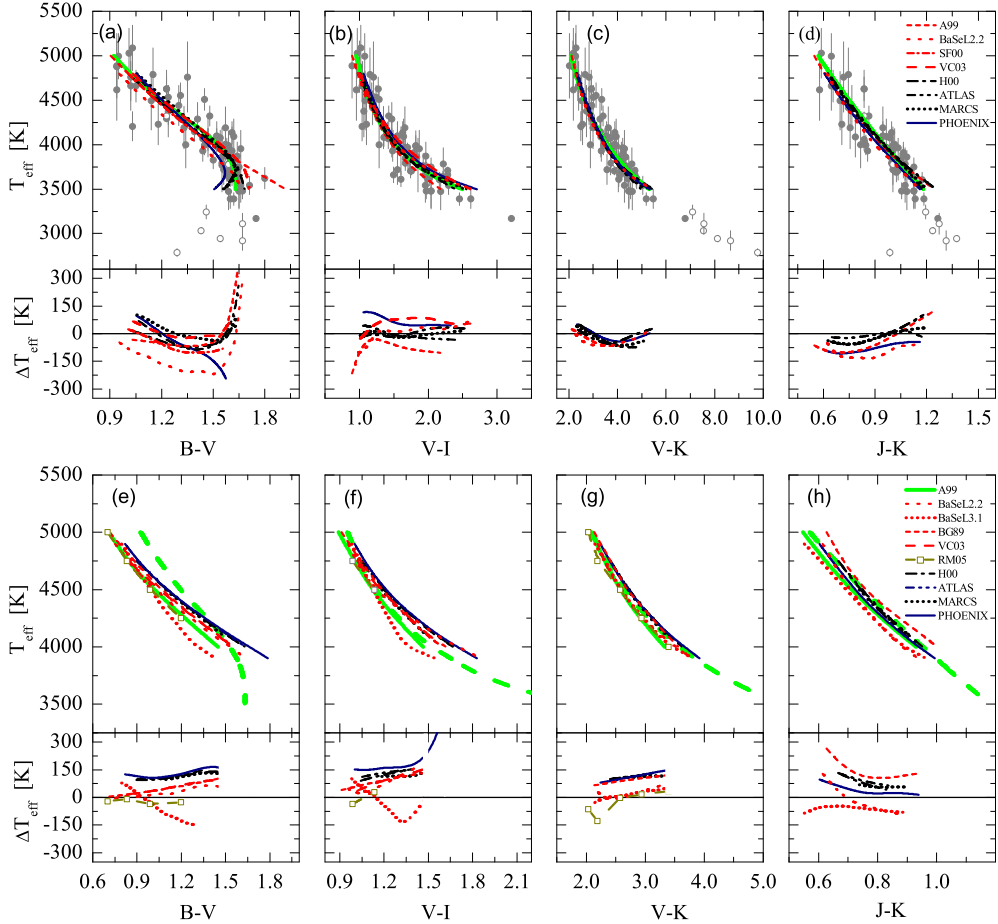}
\caption{Empirical and theoretical $T_{\rm eff}$--color relations
for late-type giants in different $T_{\rm eff}$--color planes, at
$\MoH=0.0$ (panels a--d) and $\MoH=-2.0$ (e--h). Filled circles
are late-type giants from the Solar neighborhood, variable stars
are highlighted as open circles (in both cases effective
temperatures are derived from interferometry). Thick solid line is
a best-fit to the data at $\MoH=0.0$ (panels a--d; also shown as
thick dashed line in panels e--h); thick lines in panels e--h are
$T_{\rm eff}$--color relations of A99 at $\MoH=-2.0$. Several
existing $T_{\rm eff}$--color relations are shown as well (BaSeL
3.1: Westera et al. 2002, A\&A, 381, 524; BG89: Bell \& Gustafsson
1989, MNRAS, 236, 653; RM05: Ram\'{i}rez \& Mel\'{e}ndez 2005,
ApJ, 626, 465; see \cite[Ku\v{c}inskas \etal\ 2005]{K05} for other
abbreviations), together with semi-empirical scales constructed
using synthetic colors of {\tt PHOENIX}, {\tt MARCS} and {\tt
ATLAS}. The bottom panels in each figure show the difference
between various $T_{\rm eff}$--color relations and either the
best-fit scale ($\MoH=0.0$) or $T_{\rm eff}$--color relations of
A99 ($\MoH=-2.0$), in a given $T_{\rm eff}$--color plane ($\Delta
T_{\rm eff}=T_{\rm eff}^{\rm other}-T_{\rm eff}^{\rm
bestfit/A99}$). \label{fig:TCrels} }
\end{figure}

To assess the current status in the theoretical modeling of the
spectral properties of late-type giants, we provide a comparison
of synthetic photometric colors of late-type giants (calculated
with {\tt PHOENIX}, {\tt MARCS} and {\tt ATLAS} model atmospheres)
with observations, at $\MoH=0.0$ and $-2.0$
(Fig.~\ref{fig:TCrels}). Overall, there is a good agreement
between synthetic colors and observations, and synthetic colors
and published $T_{\rm eff}$--color relations, both at $\MoH=0.0$
and $-2.0$. Deviations from the observed trends in $T_{\rm
eff}$--color planes are generally within $\pm150$\,K (or less) in
the effective temperature range of $T_{\rm eff}=3500-4800$\,K.
Synthetic colors calculated with different stellar atmosphere
models typically agree to $\sim100$\,K, within a large range of
effective temperatures and gravities. Some discrepancies are seen
in the $T_{\rm eff}$--$(B-V)$ plane below $T_{\rm eff}\sim3800$\,
K at $\MoH=0.0$, due to difficulties in reproducing the 'turn-off'
to the bluer colors which is seen in the observed data at $T_{\rm
eff}\sim3600$\,K. Note that at $\MoH=-2.0$ effective temperatures
given by the scale of \cite[Alonso \etal\ (1999, A99)]{A99} are
generally lower than those resulting from other $T_{\rm
eff}$--color relations based both on observed and synthetic
colors. This is clearly seen in all $T_{\rm eff}$--color planes,
with an average offset of $\sim130$\,K.

Obviously, reasonably good agreement can be achieved between
theoretical predictions and observed properties of late-type
giants at the level of about $\pm150$\,K; however, systematic
differences in individual $T_{\rm eff}$--color planes may easily
reach (or even exceed) $\pm100$\,K. While this may point towards
an interplay of various factors to be clarified in a dedicated
analysis (inadequacies in current theoretical models, intrinsic
differences in the atmospheres of individual stars), $\pm100$\,K
may represent a reasonable lower error margin in $T_{\rm eff}$ of
late-type giants obtainable with currently available stellar
atmosphere models.

\end{document}